\documentstyle[aps,prl, preprint]{revtex}
\begin{document}
\title{Density Matrix Renormalization Group Applied to the Ground State of the
XY-Spin-Peierls System.}
\author{L.G. Caron and S. Moukouri}
\address{Centre de recherche en physique du solide and D\'epartement de physique,\\
Universit\'e de Sherbrooke, Sherbrooke, QC, Canada J1K 2R1.}
\date{12 January 1996}
\maketitle

\begin{abstract}
We use the density matrix renormalization group (DMRG) to map out the ground
state of a XY-spin chain coupled to dispersionless phonons of frequency $%
\omega $. We confirm the existence of a critical spin-phonon coupling $%
\alpha _c\propto \omega ^{0.7}$ for the onset of the spin gap bearing the
signature of a Kosterlitz-Thouless transition. We also observe a
classical-quantum crossover when the spin-Peierls gap $\Delta $ is of order $%
\omega $. In the classical regime, $\Delta >\omega $, the mean-field
parameters are strongly renormalized by non-adiabatic corrections. This is
the first application of the DMRG to phonons.
\end{abstract}


The spin-Peierls state\cite{Bray} in quasi-one-dimensional materials has
attracted renewed attention lately because of its discovery in the organic
series\cite{Liu}\cite{Dumoulin} $(BCPTTF)_2X$ , with $(X=AsF_6,PF_6)$, and
in the cuprate compound \cite{Pouget} $CuGeO_3$. Our own interest in this
field stems from earlier findings on non-adiabatic effects or quantum
lattice fluctuations on the transition temperature $T_c$ of the closely
related (through the Wigner-Jordan transformation\cite{Bray})
electron-phonon problem. These generally tend to decrease $T_c$ \cite{CB3}%
\cite{McKenzie} and the order parameter\cite{McKenzie}\cite{Psaltakis} and
can even destroy it for spinless fermions when the phonon frequency gets
appreciably larger than the gap\cite{CB1}. The existence of this
adiabatic-quantum crossover has thus far remained only a theoretical
concept. Moreover, there are predictions of a Kosterlitz-Thouless (K-T)
transition in the quantum regime\cite{Hirsch+Fradkin} which yet need to be
confirmed. With the advent of the density matrix renormalization group (DMRG)%
\cite{White}, numerical calculations of the ground state properties of large
scale systems has now become a reality. Although phonons have not yet been
approached by this method, we will show the feasibility of such calculations.

We propose to study a XY-spin chain whose magnetic interaction depends on
the bond length. The reason is twofold. First, the {\it uniform} spin chain
enjoys an exact solution. It can be mapped into a half-filled {\it %
non-interacting} spinless fermions (pseudo-electrons) chain of width $2J$
using the Wigner-Jordan transformation\cite{Bray}. Second, the model
contains the essential elements for a spin-Peierls transition, that is
coupling to inter-molecular motion. We have chosen a dispersionless
vibrational spectra for the molecular motion along the chain direction. This
is quite acceptable as it is well known that it is the phonons at the zone
edge that couple to the spins. This also circumvents the contribution of the
hydrodynamic modes to the molecular displacements which increase
logarithmically with chain length. This is especially annoying if one wishes
to study the thermodynamic limit of the system as one has to go to long
chain lengths. We start from the following Hamiltonian 
\[
H=\sum_\ell \left\{ \omega b_\ell ^{\dagger }b_\ell +\left[ J+\alpha \left(
b_\ell ^{\dagger }+b_\ell -b_{\ell +1}^{\dagger }-b_{\ell +1}\right) \right]
\left( S_\ell ^XS_{\ell +1}^X+S_\ell ^YS_{\ell +1}^Y\right) \right\} 
\]
in which $\omega $ is the vibrational energy (or frequency as $\hbar =1$
throughout), $b_\ell $ $(b_\ell ^{\dagger })$ is the annihilation (creation)
operator for a vibration on molecule $\ell $, $J$ is the magnetic
interaction, $\alpha =g/\sqrt{2\omega }$ where $g=\left( \partial J_{\ell
,\ell +1}/\partial \ell \right) $ is the standard spin-phonon interaction in
the classical paper of Bray {\it et al.}\cite{Bray}, and ${\bf S}_\ell $ is
the local XY spin of value $\frac 12$.

The expected crossover and the nature of the different regimes warrants
further examination at this time. In the two-cutoff analysis\cite{CB1}, the
natural energy cutoff for the pseudo-electrons is the bandwidth for
characteristic energies (e.g. the gap $\Delta $) larger than $\omega $. This
is the classical regime in which the adiabatic approximation prevails. But
for characteristic energies less than $\omega $, there is a new cutoff $%
\omega $ and the pseudo-electron-phonon coupling generates instantaneous
backward-scattering $g_1\approx -8\alpha ^2/(\omega \pi {\rm v}_{{\rm F}})$
and Umklapp $g_3\approx -g_1$ couplings (in the renormalization group
notation\cite{Voit}\cite{CB2}\cite{Caron}), where ${\rm v}_{{\rm F}}$ is the
Fermi velocity of the pseudo-electrons. This is the quantum regime in which
the interactions are unretarded and quantum fluctuations impregnate the
ground state. It turns out that, for spinless fermions, the local character
of the backward scattering cancels out. It is only the non-local character
(wave number dependence) that can give rise to non-trivial effects. The
renormalization equations\cite{Dumoulin}\cite{Black} show that {\it unless}
the non-local contribution to $g_3$ has the right sign and has a bare
(initial) value larger than a certain threshold, the Umklapp processes are
irrelevant ($g_3$ renormalizes to zero) and the quantum system is gapless.
If the threshold condition is met, the Umklapp processes and the vertex
function grow to infinity. This signals the onset of a gap in the
excitations. The transition is believed to be of the K-T type\cite{den Nijs}%
\cite{Hirsch+Fradkin}. We thus propose to look for a threshold value $\alpha
_c$ for the appearance of a gap and for the signature of the
Kosterlitz-Thouless transition. The formula derived by Baxter\cite{Baxter}
for the order parameter $P$ is of the form $P\propto \lambda ^{-1}\exp
\left( -\lambda ^{-1}\right) $ where is $\lambda \propto \left( T_c-T\right)
^{1/2}$ for a thermodynamic transition. For a ground state transition,
identifying $T$ with $g_1\propto \alpha ^2$ suggests $\lambda \propto \left(
\alpha ^2-\alpha _c^2\right) ^{1/2}$. We shall refer to this last form as
the effective K-T coupling.

We now turn to the use of the DMRG. Its inherent difficulty in dealing with
vibrations is the infinite dimension of the Hilbert space for phonons. The
most direct way around this problem is to truncate the space. This is
obviously unsatisfactory whenever the average molecular displacements
involve too many virtual phonons $n_\ell =\left| \left\langle b_\ell
\right\rangle \right| ^2$. A crude estimate for this can be obtained by
equating the increase in elastic energy to half the decrease in electronic
energy $U$. For single electrons, as in the polaron problem, one gets is $%
n_\ell \approx U/2\omega $ which obviously precludes any serious study of
the deep adiabatic region $\omega \ll U$\cite{Marsiglio} with a limited
phonon space. In the case of a collective state, such as the spin-Peierls
(or Peierls) modulated state, the same analysis for non-interacting
pseudo-electrons leads, in a mean-field type analysis, to the criterion $%
n_\ell \approx \Delta ^2\ln (J/\Delta )/2\omega \pi J$. One gains a factor $%
\Delta \ln (J/\Delta )/\pi J$ over the single-electron situation, allowing a
deeper incursion into the adiabatic region $\omega \ll \Delta $. In our
case, we typically kept $n_\ell $ less than 3 and $\omega $ was as small as $%
\Delta /25$ under the best conditions.

In the DMRG procedure a set of internal sites are coupled, via the
inter-site part of the Hamiltonian, to environmental blocks. The DMRG
proposes an iteration algorithm for the growth of the environmental blocks
and thus, for the size of the system. Our adaptation of the method, for each
set of parameters $\left( \alpha /J,\omega /J\right) $, was the following.
At each iteration, we considered a single site coupled to the environment
blocks generated by the previous iteration. This central site had two spin
states and $M_v$ vibration states, equal to anywhere from three (at small
gaps) to twenty (for the larger values of $\alpha $). $M_v$ could be
estimated by requiring that the local vibration subspace properly describe a
coherent state of $n_v$ phonons, that is with an error much less than the
allowed DMRG truncation error. The total number of sites in the chain was
chosen to be even such that, with open boundary conditions, the chain would
relax to a unique broken-symmetry state. Consequently, the number of sites
of the bordering environment blocks differed by one, there being one long
block and a one-site shorter block. We kept $M_b$ state in each block, a
number varying from 60 to 120 (for the smaller gaps), and the corresponding
matrix elements of all operators coupling the central site to the bordering
sites of each block. We targeted the four lowest energy states in the
superblock using the Davidson-Liu\cite{Davidson} algorithm, two in each of
the total spin subspaces having $S^Z=0,1$. This way, we were able to
reliably access the spin gap energy on the ansatz that, in the spin-Peierls
state, the ground state is a singlet $S=0$ state and the spin gap is to an $%
S=1$ state. These four target states were then used to project out the
density matrix $\rho _{ij}$ of the two new blocks formed by the direct
product of the spaces of the left (right) block and the added site. The z
component of the spin being a good quantum number, we used it to fragment
the block spaces into more manageable sizes. The density matrix was
diagonalized and the $M_b$ highest energy eigenstates were kept. The
truncation error $Tr\ \rho -\sum_{i=1}^{M_b}\rho _{ii}$, which was used as a
minimal error indicator, was kept at a few times $10^{-5}$ or less. The
block sizes thus increased by one and the chain length, by two sites at each
iteration. The total length varied from 100 to 600 sites at the smaller
gaps. At the longer chain lengths, the numerical accuracy on the gap greatly
suffered and finite size scaling arguments were used to analyze the data.
The error on the gap is estimated to be 10\% under these strained conditions.

Fig. 1 shows typical values of the gap plotted in such a fashion as to bring
out the characteristic K-T form $\Delta \propto (\alpha ^2-\alpha
_c^2)^{-0.5}\exp \left( -b(\alpha ^2-\alpha _c^2)^{-0.5}\right) $. The
signature is clear and there is indeed a critical spin-phonon
(pseudo-electron-phonon) coupling $\alpha _c$ for the onset or long-range
order. Its values are obtained by least-squares fitting and are plotted in
Fig. 2. This phase diagram identifies, for the first time, the existence of
a power-law relationship $\alpha _c\sim \omega ^{0.7}$. The average
molecular displacement $u=\left| \left\langle b_\ell ^{\dagger }+b_\ell
\right\rangle \right| $ also seems to obey a similar law although the fit is
less reliable as $u$ varies much more slowly. Fig. 3 hints to the probable
existence of a power relationship between $\Delta $ and $u$ which one should
expect if both parameters follow the K-T formula. We could not define one
unambiguously.

Fig. 3 points to the identification of a crossover between two regimes. One,
on the left of Fig. 3, is characterized by a rapid increase of $\Delta
/\alpha u$ and the other one, on the right, shows a rather flat dependence $%
\Delta /\alpha u\sim 1$. As a matter of comparison, the mean-field solution%
\cite{Bray} predicts $\Delta /\alpha u=2$. It is thus tempting to identify
this latter region, which has a mean-field like behavior, as the classical
or adiabatic regime described earlier. Fig. 4 confirms this assignment. It
shows a crossover from a classical region, with a gap of the mean-field form 
$\Delta =\Delta _o\exp (-AJ\omega /\alpha ^2)$, into another regime, the K-T
one of Fig. 1. Moreover, this crossover occurs at the expected value $\Delta
\approx \omega $ (at the position of the arrows in Fig. 4). This crossover
is also shown in the phase diagram (Fig. 2). The classical region is
obviously limited to $\Delta <\omega $. We have found the gap to saturate at
a value very near $J$ for large $\omega $. This is why the adiabatic regime
is restricted to $\omega <J$ in Fig. 4. The gap prefactor $\Delta _o$ and
the leading coefficient $A$ in the exponential are strongly frequency
dependent as shown in Fig. 5. We find that $A$ extrapolates to 0.44 and $%
\Delta _o$ to $1.3J$ at zero frequency (calculations cannot be performed at $%
\omega \rightarrow 0$ since the number of virtual phonons goes to infinity).
These values are close to the expected mean-field ones of $\pi /8$ and $4J/e$%
, respectively\cite{Bray}\cite{Hirsch+Fradkin}. The difference is
significant, however, in view of a least-squares error of order 2\% for
these parameters. It is perhaps our linear extrapolation from finite
frequencies which is improper. This conclusion is borne out by the plot in
Fig. 5 of the plateau value of $\Delta /\alpha u$ in Fig. 3 (an average
value in the classical region). It clearly shows that the classical regime
does not satisfy the mean-field condition\cite{Bray} $\Delta /\alpha u=2$
and that the extrapolation to $\omega \rightarrow 0$ is surely far from
linear. It is however likely that this ratio can reach the value of 2 and
that our results are consistent with the mean-field limit at $\omega =0$.
The parameters $\Delta _o$ and $A$ in the classical regime expression for
the gap are seen in Fig. 5 to be strongly renormalized by non-adiabatic
phonon contributions. Moreover, there seems to be a crossover, within the
classical regime, between two behaviors. For $\omega \leq .05J$, the
renormalization of $\Delta _o$ and $A$ are identical. This suggests a common
origin to the renormalization, i.e. that it is the magnetic interaction $J$
which changes at these smaller frequencies. For $J>\omega >.05J$, these two
parameters evolve differently, the extra renormalization of $A$ coming
seemingly from vertex corrections to $\alpha $. The reason for this peculiar
behavior is unclear.

We were also able to monitor the change in phonon frequency in the classical
regime. We find, as Wu {\it et al}.\cite{Wu} did, that the renormalized
phonon frequency $\omega _r$ decreases as $\Lambda =8\alpha ^2/\pi \omega J$
does. For instance, we find at the crossover $\omega /\Delta =1$ that $%
\omega _r/\omega \approx 0.5,0.7$ for $\omega =.025,0.3$ or $\Lambda \approx
0.3,1.3$. This renormalization is stronger than the one Wu {\it et al}. found
for spin $1/2$ fermions.

In conclusion, we have not only verified the predictions of a
Kosterlitz-Thouless transition and of a classical-quantum crossover in the
spinless-fermion-phonon problem but we have also been able to produce a
quantitative phase diagram of the spin-Peierls system and deduce a power law
dependence of the critical spin-phonon coupling on the phonon frequency . We
have also shown that the DMRG can be used advantageously on phonon problems
be they only dispersionless in this paper.

We wish to thank S. White for his suggestions on more efficient computer
memory management. Also many thanks to our colleagues C. Bourbonnais and
A.-M. Tremblay for many long discussions of our results. We finally
acknowledge the financial support of the Natural Sciences and Engineering
Research Council of Canada and the Fonds pour la Formation de Chercheurs et
d'Aide \`a la Recherche of the Qu\'ebec government.

\figure{FIG. 1. Semi-log plot of the spin gap }$\Delta ${\ as a function of
the reciprocal of the effective K-T coupling, for two values of the phonon
frequency. The lines, as well as the quoted values of }$\alpha _c,${\ are
least squares fits to the formula of Baxter in the quantum region.}

\figure{FIG. 2. Phase diagram of the system. The full line is a fit of the
Kosterlitz-Thouless transition line. It has the power-law dependence $\alpha
_c^{-2}\propto \omega ^{-1.4}$. The dashed line straddles the
classical-quantum crossover }$\Delta \approx \omega ${\ and ends at }$\omega
\approx J${.}

\figure{FIG. 3. Ratio of the spin gap }$\Delta ${\ to the lattice modulation
amplitude }$u$ {times the spin-phonon coupling }$\alpha ${\ as a function of
the square of the effective K-T coupling, for two values of the phonon
frequency. The arrows point to the axis each data set refers to.}

\figure{FIG. 4. Semi-log plot of the spin gap }$\Delta ${\ as a function of
the reciprocal square of the spin-phonon coupling }$\alpha ${, for two
values of the phonon frequency. The straight lines are least squares fits to
the mean-field like expression in the classical regime. The arrows point to
the axis each data set refers to and also indicate where the
classical-quantum crossover occurs.}

\figure{FIG. 5. Evolution of the parameters $\Delta _o$ and $A$ of the
mean-field type expression for the spin gap }$\Delta ${\ as a function of
frequency }$\omega ${, normalized to their zero frequency values }$\Delta
_o(0)$ and $A(0)${\ which were linearly extrapolated from the data at finite
frequencies. Also shown is the frequency dependence of the average value of }%
$\Delta _o(\omega )/\alpha u$ in the classical region, where $\alpha $ {the
spin-phonon coupling and }$u$ is the average molecular displacement. This
latter ratio should equal $2$ in a mean-field solution. {The arrows point to
the axis each data set refers to.}

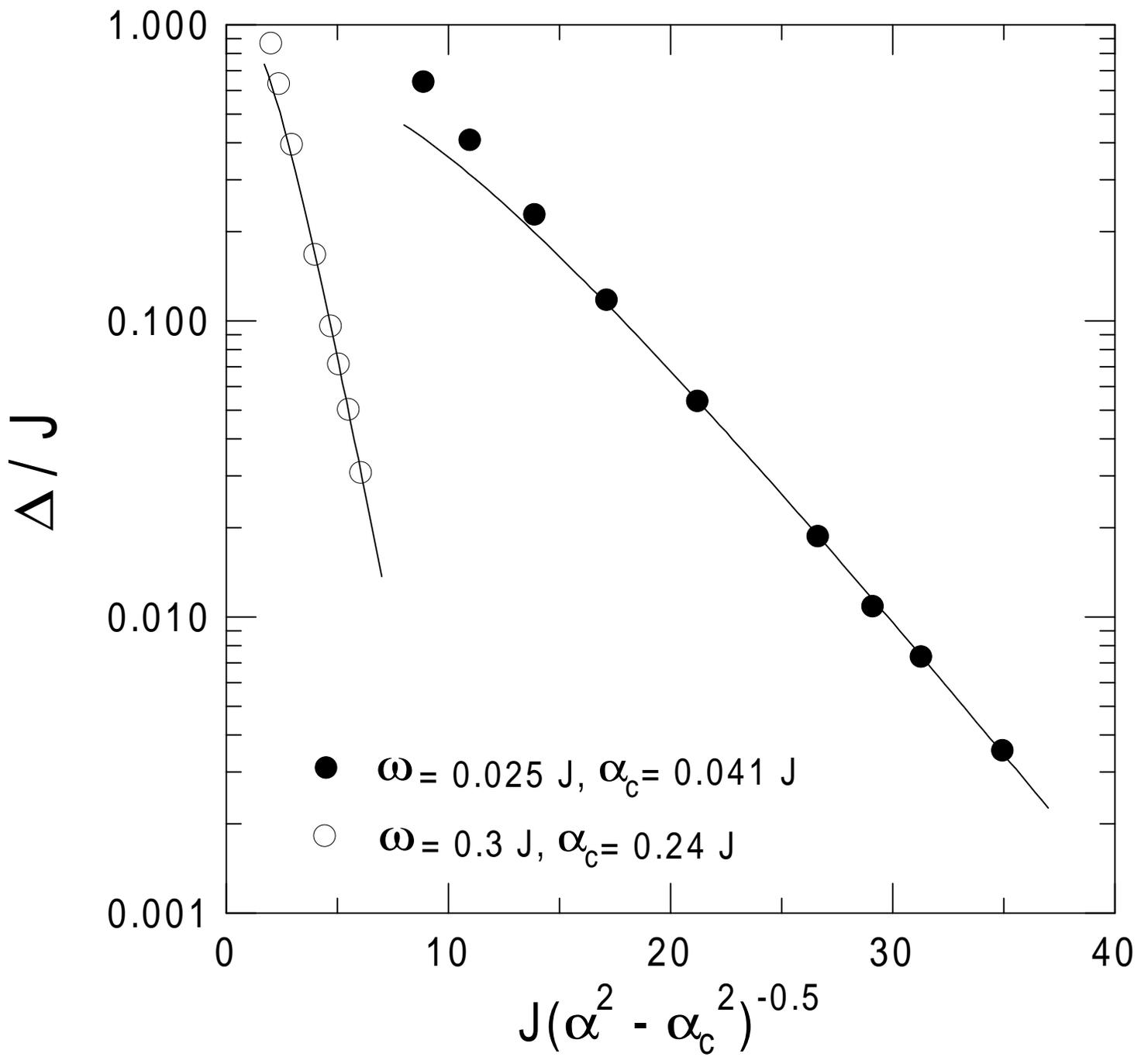
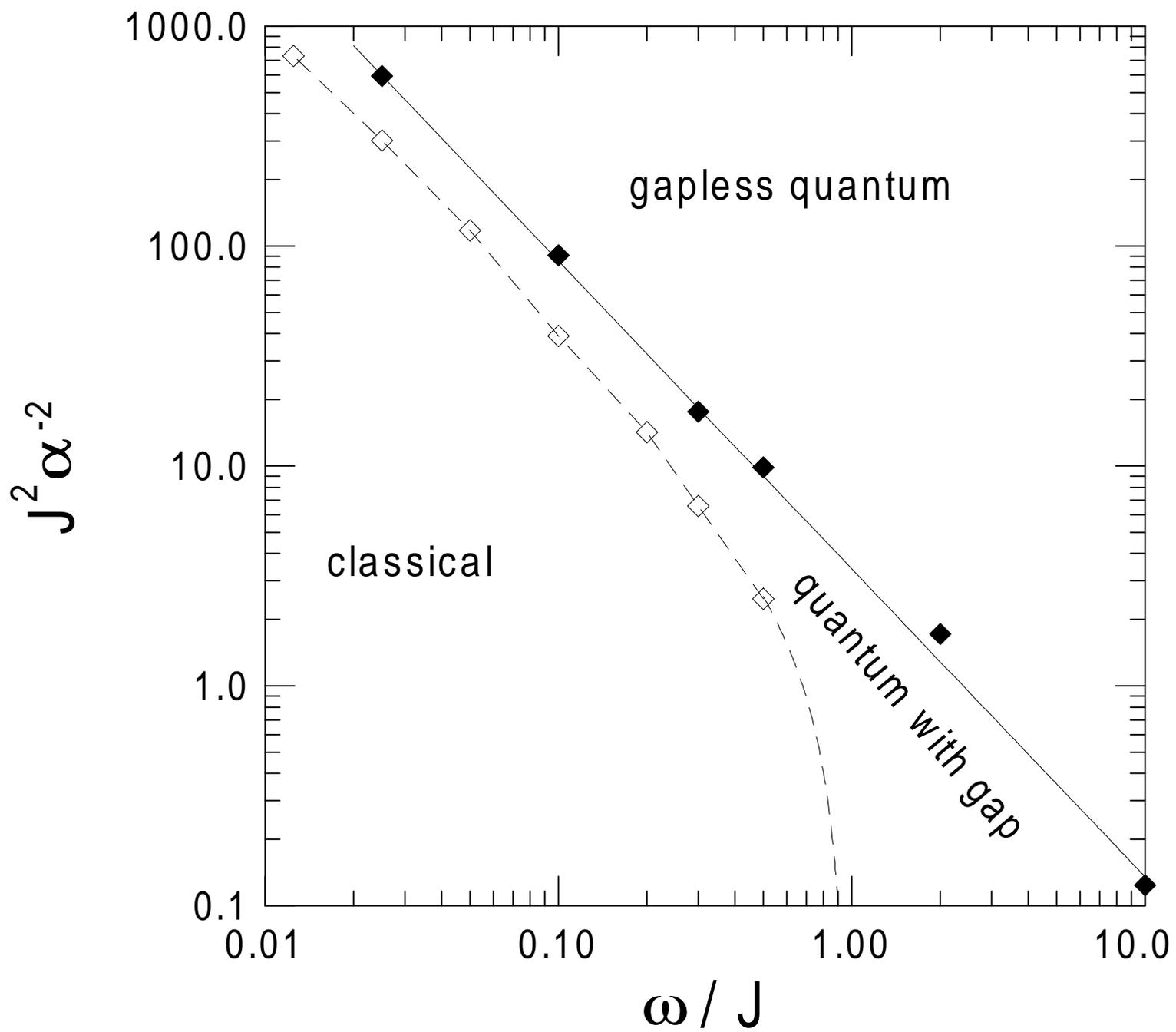
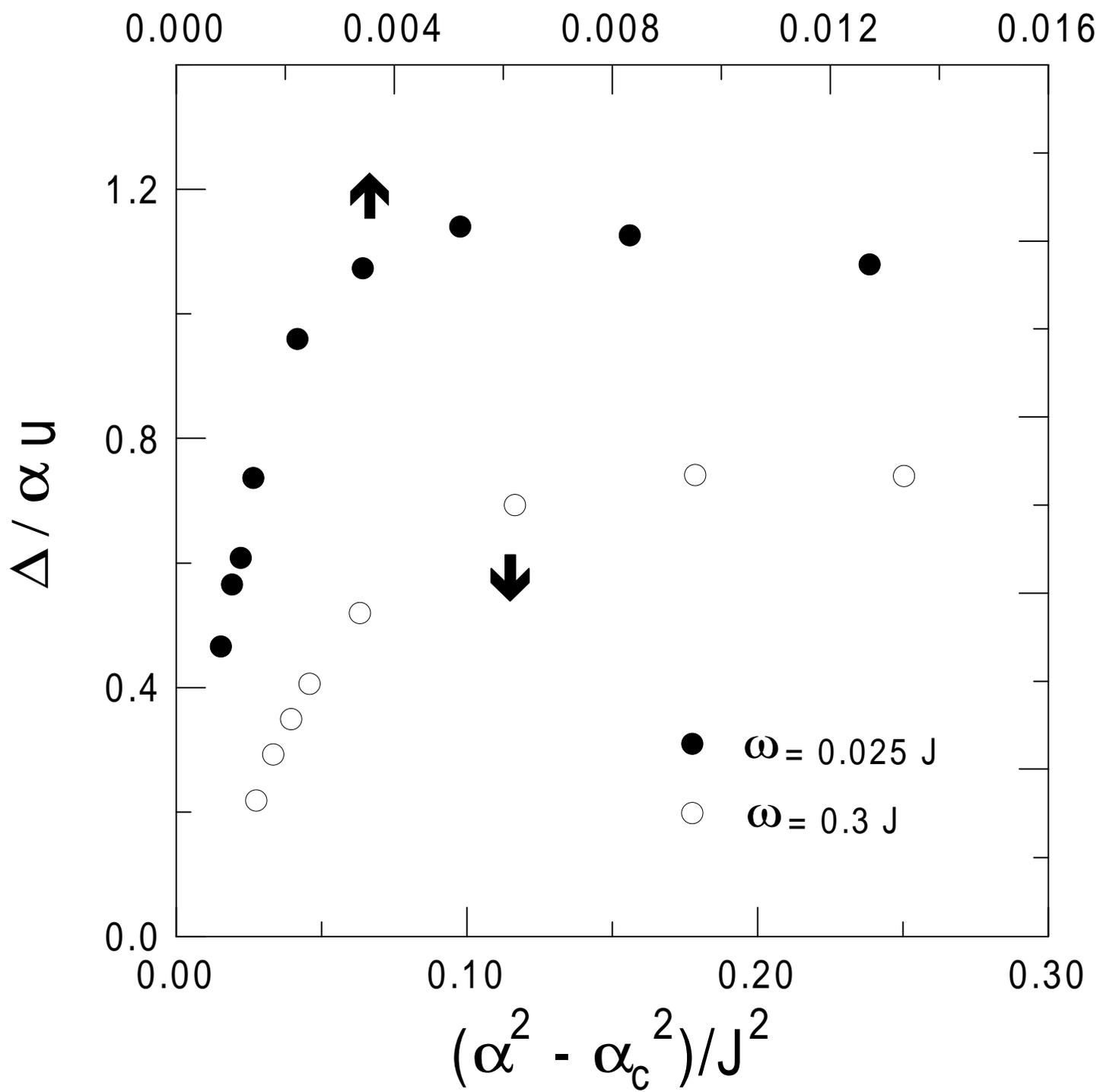
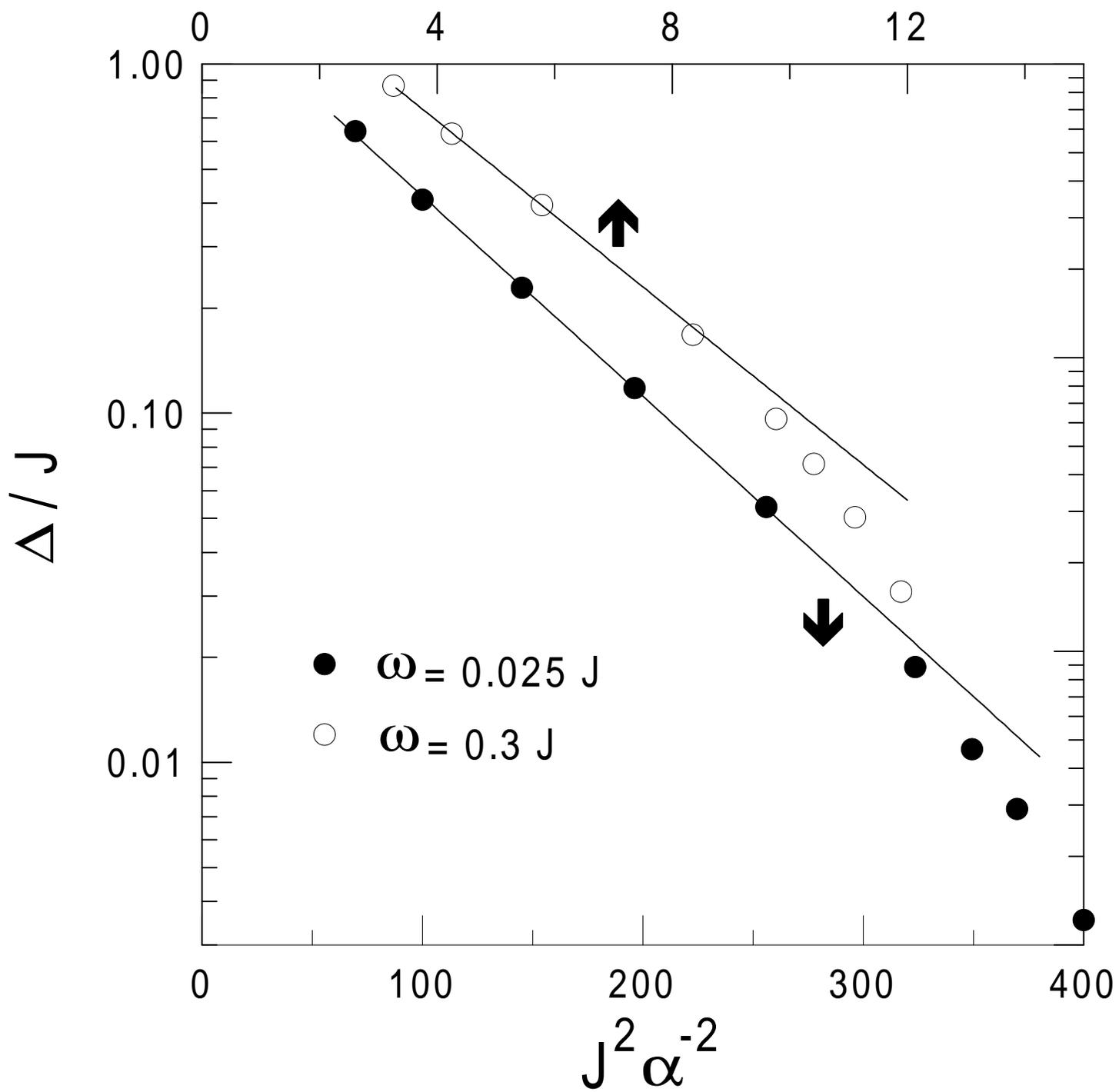
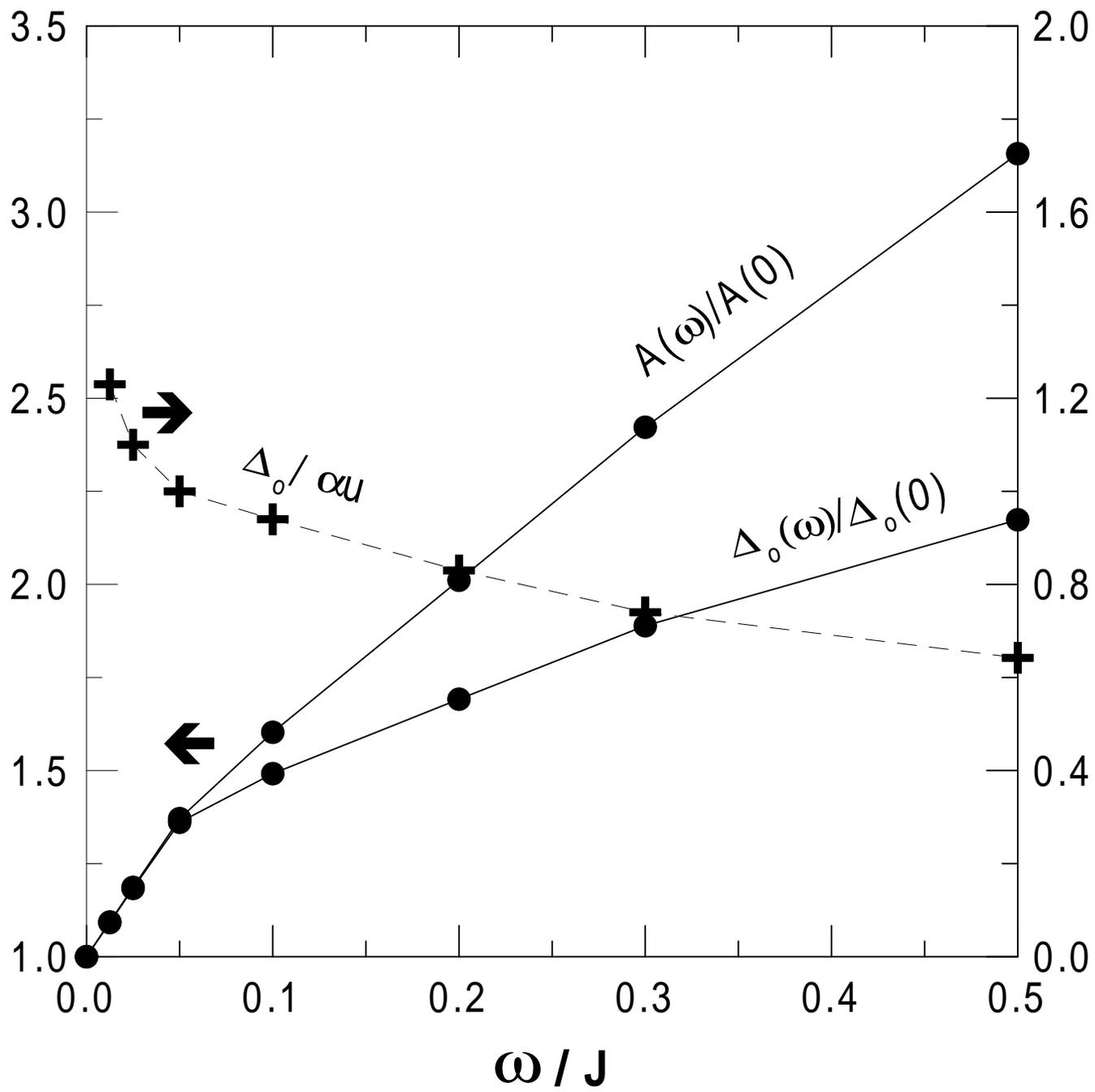
\end{document}